\documentclass{article}
 \usepackage{amsmath}
 \usepackage{amssymb}

\usepackage{geometry}
\usepackage{authblk}

\usepackage{graphicx}
\usepackage{dcolumn}
\usepackage{bm}
\usepackage{hyperref}
\usepackage{xcolor}
\usepackage[utf8]{inputenc}
\usepackage{newunicodechar}
\newunicodechar{̂}{\^{} }
\usepackage[T1]{fontenc}

\usepackage[backend=biber, style=numeric, sorting=none]{biblatex}
\bibliography{biblio}

\begin{document}

\title{Effective Quantum Gravitational Collapse in Metric Variables:\\ The $\bar{\mu}$ Scheme}

\author[1]{Lorenzo Boldorini\thanks{boldorini.1843532@studenti.uniroma1.it}}
\author[1,2]{Giovanni Montani\thanks{giovanni.montani@enea.it}}
\affil[1]{Department of Physics, “La Sapienza” University of Rome,\\ P.le Aldo Moro 5, 00185 Rome, Italy}
\affil[2]{ENEA, Fusion and Nuclear Safety Department, C. R. Frascati,\\Via E. Fermi 45, 00044 Frascati (RM), Italy}

\date{\today}

\maketitle

\begin{abstract}
We study, using the metric variables, how an effective theory for the Oppenheimer-Snyder gravitational collapse can be built with the $\bar{\mu}$ scheme from Loop Quantum Gravity (LQG). The collapse is analyzed for both the flat and spherical models. In both scenarios the effective theory make possible to avoid the formation of the singularity. The source of this is found in the presence of a negative pressure term inside the stress-energy tensor of the gravitational field. This pressure is analyzed and is concluded that the effective polymer model is the reason why the negative pressure appears. A characterization of the solutions for both models is also carried out, showing that the collapse is altered and avoided in favor of a transition from a black hole state to a white hole one, transition that occurs when the collapse has reached a Planckian regime.
\end{abstract}

\newpage
\section{\label{sec:introduction}INTRODUCTION}
Black holes are one of the most discussed aspects of the theory of General Relativity (GR), since all black holes contain at their core the singularity \cite{Misner:1973prb} \cite{Weinberg2008}, a point that possess infinite curvature, signaling that GR has reached its domain of applicability.\\
The theory predicts that such objects are generated in nature through the means of gravitational collapse, that is, when a star becomes too massive for its internal pressure to counteract the gravitational pull generated by its mass.\\
To prevent the formation of such singularities one is led to build theories that encompass GR, but that also include quantum effects that could prevent the breakdown of the theory.\\ 
One of the first and most conservative framework considered for this job is the Wheeler-DeWitt quantization of the gravitational field \cite{WdW67}, a canonical quantization of the metric. This framework was soon changed and evolved into what is now known as the Loop Quantum Gravity (LQG) theory, a canonical quantization scheme based on the Ashtekar-Barbero-Immirzi’s connection variables. This approach mimics the same canonical quantization procedure applied to gauge theories, of which gravity is a special type.\\ Being different from gauge theories usually encountered in high energy physics, due to its diffeomorphism invariance, there are key differences with the usual quantization. A comprehensive description of the theory can be found at \cite{Thiemann_2007}.\\
Since the development of the theory a lot of work was done in establishing a connection with observations. In particular a lot of attention was devoted to the physics of gravitational collapse, specifically the Oppenheimer-Snyder (OS) model \cite{Oppenheimer_Snyder_1939} of gravitational collapse, a model describing the collapse of pressureless dust.\\
The key features reported in the literature regarding the quantization of collapse models are divided mainly in models that predict a bouncing behaviour, found at \cite{Rovelli_2018, DeLorenzo2016, Husain2022, Achour2020v1, BenAchour2020, Barca2023, Bambi2013, LTB_Kief, Liu2014, bobula23, lewandoski23}, and models that predict the formation of shockwaves, found at \cite{Fazzini2023, Fazzini2024, HusainRev2022, Cipriani2024, gieselliu24}.\\
The bouncing models predict then the formation of white holes due to quantum transitions, while the shockwave models predict that the formed shockwave ultimately dissolve the event horizon of the black hole generated during the collapse.\\
Both results avoid the formation of the singularity during the collapse due to the introduction of quantum effects.\\
Our scope in this work will be to adopt the OS model for gravitational collapse and make use of an effective framework used in the LQG community to build an effective description of the collapse in metric variables.\\
The scheme is referred to as the $\bar{\mu}$ scheme \cite{muscheme}  and serve the purpose of introducing and accounting for the discreteness of the area present in the LQG theory \cite{Rovelli_1994} inside the Hamiltonian of the theory, through a polymerization of the model \cite{Corichi_2007}.\\
\textcolor{black}{Polymer models, or schemes, are a way to introduce in the classical theory quantum corrections from the full theory. Those corrections come in LQG from the fact that the connection is not well defined, but its holonomy is, so that one polymerizes the connection, which is recovered from the holonomy in the limit of vanishing polymerization. The lattice step used to polymerize the connection usually comes from the discrete spectrum of the area operator. In this manuscript we will adopt the $\bar{\mu}$ scheme, as already introduced earlier.\\}
We will build an effective Hamiltonian for the OS model with the aim of considering the trajectories to be those of the expectation values of the quantum operators,  peaked around the classical trajectory.\\
Our aim in this paper is to understand what are the effects of the $\bar{\mu}$ scheme on the classical dynamics, namely if the singularity in the metric is removed and how this compares with the previously analyzed scheme in the metric variables \cite{Boldorini_2024}.\\
\textcolor{black}{The reason behind this choice lies in the fact that quantization of different variables result in different theories. One of the aims of this manuscript is thus to understand whether the bouncing dynamics in polymerized models of LQG is exclusive to connection variables or rather it's a generic feature of polymerized effective theories, regardless of the variables taken in consideration.}
The main results of the following paper show that the singularity is always removed in favor of a bounce, hinting at a transition from a black hole solution to a white hole one, that happens always in the Planck regime.\\
We also show that such effective model can be understood as the introduction of a negative pressure inside the stress-energy tensor of the gravitational field itself.\\
\textcolor{black}{Furthermore, a detailed analysis of the junction conditions with the effective model is performed: a surface membrane of stress-energy is found, and violations of the energy conditions of General Relativity are studied. It is shown that the effective model breaks down before reaching the minimal surface, due to the star reaching the region of strong quantum effects.}\\
The paper is organized as follows. In section \ref{sec:clas_OS} we will briefly introduce the classical Hamiltonian model describing the Oppenheimer-Snyder collapse, along with the definition of some useful parameters used throughout this manuscript.\\
In section \ref{sec:effective_model} we introduce the effective model, encompassing the improved polymer scheme. Right after we proceed to solve the equations of motion for both the
area and momentum, highlighting after each solution the key features.\\
In section \ref{sec:phenom} we present the phenomenology of the new effective model, showcasing the differences with the previously adopted model.\\
The main differences will rely in the regime where the quantum effects take place and alter the classical behavior of the system. We also analyze the junction conditions for the effective model, finding a scale for which the effective model breaks down.\\
At last we close the paper in section \ref{sec:conc} with a summary of the analysis performed.\\
We will use geometrized $(G=c=1)$ units in the following discussion.

\section{\label{sec:clas_OS}THE CLASSICAL OPPENHEIMER-SNYDER MODEL}
In this section we are going to recall the gravitational collapse model developed in \cite{Boldorini_2024}.\\
The adopted model is the solution found by Oppenheimer and Snyder \cite{Oppenheimer_Snyder_1939}, reformulated using ADM metric variables \cite{ADM_1959} \cite{Poisson_2004}, in detail the description of the collapse will be given  in terms of the surface area of the collapsing ball of dust and its conjugate momentum. \\
Denoting as $\left(t,r,\theta, \phi\right)$ the spherical coordinates describing the Oppenheimer-Snyder solution, the model is given by a spherical ball of dust with its outermost layer at a fixed coordinate radius $r=r_S$. Those coordinates have only the role of parameters for the canonical variables, and serve no purpose for the dynamics of the system, which is described only by the canonical variables.\\
The dynamics of the dust is described by the Brown and Kuchař \cite{Brown_Kuchar_1995} canonical variables $\tau$, its proper time, and $P_\tau$, its proper energy, while the dynamics of the gravitational field is described by the metric field $\bm{g}$: for the spacetime portion inside the dust ball it is the cosmological Friedmann-Lemaître-Robertson-Walker (FLRW) metric \cite{Misner:1973prb}, while the portion outside the ball is described by the Schwarzschild metric\cite{Misner:1973prb}.\\
The two metrics are joined at the dust ball surface with the Israel junction conditions \cite{Israel_1966}.\textcolor{black}{We'll return to the analysis of such conditions later in subsection \ref{subsec:junction}, as effective theories will deviate from the classical behavior.}\\
The canonical variables that describe the interior are $\mathcal{A}$, the proper surface area of a sphere of proper radius $R_s$, and $P_{\mathcal{A}}$, its conjugate momentum. \\
\textcolor{black}{Those are related to the FLRW metric:
\begin{equation}
\label{eqn:FLRW_ds}
    ds_-^2 = -N_\tau(t)^2dt^2 + \frac{a(t)^2}{1-\epsilon r^2}dr^2 + a(t)^2r^2 d\Omega^2
\end{equation} 
By: 
\begin{equation}
    \mathcal{A} = 4\pi a(t)^2 r_S^2 = 4\pi R_s(t)^2
\end{equation}\\
The outside is described by the Schwarzschild metric parametrized by the Generalized Painlevé-Gullstrand coordinates, for which the metric reads:
\begin{equation}
\begin{split}
\label{eqn:SCH_ds}
     ds_+^2 = -F(R)dT^2 +2&\sqrt{1 - \mathcal{C}_S F(R)} dTdR + \mathcal{C}_S dR^2 + R^2 d\Omega^2\\
     &\mathcal{C}_S = \frac{1}{1-\epsilon r_S^2}
\end{split}
\end{equation}
With $T=T(t)$ and $R=R(t)$ parametrized by the coordinate time $t$.\\
When we choose $N_\tau = 1$, then coordinate time $t$ will equal the comoving time $\tau$, so the functions $T$ and $R$ will become functions of the proper time of the dust.\\
}
More details about this construction can be found in the original article \cite{Boldorini_2024} and in \cite{Kiefer_Mohaddes_2023}\cite{Schmitz_2020}\cite{Painleve_1921}\cite{Gullstrand:1922tfa}\cite{Gautreau_Hoffmann_1978}.\\
We set:
\begin{equation}
\kappa_S = \epsilon r_S^2
\end{equation}
Where $\epsilon = 0, +1$ is for the flat and spherical FLRW metrics respectively.\\
We also recall that:
\begin{equation}
M_S = \Xi P_\tau
\end{equation}
Where the $\Xi$ parameter is the rateo between the flat and spherical volume of the dust ball. \\
Defining:
\begin{equation}
   \Upsilon^\epsilon_S  =  \int_0^{r_S} dr \frac{r^2}{\sqrt{1-\epsilon r^2}} 
\end{equation}
Then the parameter $\Xi$ is given by:
\begin{equation}
\Xi = \frac{r_S^3}{3 \Upsilon^\epsilon_S }
\end{equation}
\\
The action in those variables reads:
\begin{equation}
        \label{eqn:act_SQ}
S =\int dt \left\lbrace\left(P_\tau \dot{\tau}  + P_{\mathcal{A}} \dot{\mathcal{A}} -N_\tau \tilde{\mathcal{H}}\right) 
 +  \int_{r_S}^{\infty}dr\left[\left(\tilde{P}_R\dot{R} + P_T\dot{T}\right) - \left(\beta^T P_T +\beta^R  \tilde{P}_R\right)\right] \right\rbrace  
\end{equation}
\\
With the Hamiltonian is given by:
\begin{equation}
\label{eqn:ham_SQ}
 \begin{split}
&\tilde{H}= N_\tau \tilde{\mathcal{H}} \\ 
&\tilde{\mathcal{H}} = -\frac{\Xi }{2}\left(32\pi^{\frac{3}{2}}P_{\mathcal{A}} ^2\sqrt{\mathcal{A}} + \frac{\kappa_S }{\Xi ^2} \frac{\sqrt{\mathcal{A}}}{\sqrt{4\pi}}  -2 \frac{P_\tau }{\Xi }\right) 
  \end{split}  
\end{equation}
\\
The theory possesses the following constraints, coming from the equations of motion for $N_\tau$, $\beta^T$ and $\beta^R $:
\begin{equation}
\label{eqn:cosfin}
\tilde{\mathcal{H}} \approx 0 \qquad P_T \approx 0 \qquad \tilde{P}_R \approx 0
\end{equation}
\\
The last two equations in (\ref{eqn:cosfin}) are such that the variables $R$ and $T$ are not dynamical: this means that the exterior geometry is fixed, up to the diffeomorphisms generated by $ P_T$ and $ \tilde{P}_R$.\\
From now on we will work with the choice of $N_\tau = 1$, the comoving gauge, that give rise to the following Hamilton equations:
\begin{align}
			\dot{\mathcal{A}} &= -32 \pi ^{3/2} \Xi  P_{\mathcal{A}} \sqrt{\mathcal{A}}\label{eqn:A_clas_eom}\\
			\dot{P}_{\mathcal{A}}&= \frac{\kappa_S }{8\Xi    \sqrt{\pi\mathcal{A}}} +\frac{8\Xi \pi ^{3/2} P_{\mathcal{A}}^2}{\sqrt{\mathcal{A}}}\label{eqn:PA_clas_eom}\\
			\dot{\tau}  &= 1\label{eqn:t_clas_eom}\\
			\dot{P}_\tau   &= 0\label{eqn:Pt_clas_eom}
\end{align}
Those will be helpful later in section \ref{sec:phenom} when deriving the pressure associated to quantum effects.

\section{\label{sec:effective_model}THE EFFECTIVE MODEL}
This section will be centered on the construction of a new improved effective dynamical model, where the effective model will be built to take into accounts the quantum effects introduced by a polymer quantization of the theory \cite{Corichi_2007}.\\
First we will introduce the effective model and the corresponding equations for the sphere surface and momentum. Later we will solve those equations, showing that non trivial effects will affect both solutions.\\
Let’s now introduce the polymer quantization scheme.
\subsection{\label{subsec:poly}The  $\bar{\mu}$ Effective Model}
The effective theory is built using the technique called polymer quantization \cite{Corichi_2007}. \\
This quantization procedure start by exponentiating the two canonical variables, the coordinate $q$ and the momentum $p$, to the respective Weyl elements $U(\alpha)=e^{i\alpha q}$ and $V(\beta)=e^{i\beta p}$.\\
Then one select a representation where one of the two exponential is discontinuous. This results in one of the two operators to be undefined as a derivative operator, and thus needs to be regularized by the introduction of a lattice on the canonical coordinate to be derived.\\
Inspired by the LQG literature \cite{Rovelli_1994}\cite{Rovelli_1995}, we choose a representation where the momentum $P_\mathcal{A}$ is not defined and the area $\mathcal{A}$ is discretized.\\
In our previous work\cite{Boldorini_2024} we worked with a constant discretization of the area, known in the LQG community as the $\mu_0$ scheme\cite{Ashtekar_2006}. This discretization scheme was shown to be physically unpleasant, since it predicts Planckian effects in a macroscopic regime.\\
It was shown by Corichi and Singh\cite{Corichi_Singh_2008} that the only possible consistent scheme to polymerize the effective theory is the so called $\bar{\mu}$ scheme\cite{muscheme}, a scheme where the quantum gravity effects arise only in the Planckian regime.\\
In our work we'll follow a procedure similar to the one showcased in \cite{Giesel2023, Liu2023} to implement the $\bar{\mu}$ scheme in spherically symmetric spacetimes. Here an homogeneous cosmological model is matched to a classical exterior geometry, by the use of matching conditions. \\
In metric variables, the  $\bar{\mu}$ scheme is related to a constant polymerization of the volume\cite{barca_montani_giovannetti}, defined as $\mathcal{V} = \frac{4\pi}{3}R_s^3$.\\
\textcolor{black}{Since we are polymerizing only the interior FLRW metric, as it is the only dynamical part of our system, the choice of a constant volume seems the most adeguate, as such a geometry is isotropic.}\\
Defining the lattice step in the volume variable $\Delta_\mathcal{V}$, that plays the role of minimal volume, to be constant, we see that it implies that the lattice step on the area variable must be:
\begin{equation}
\Delta_\mathcal{A} = \frac{3\Delta_\mathcal{V}}{R_s} =  \frac{6\sqrt{\pi}\Delta_\mathcal{V}}{\sqrt{\mathcal{A} }}.
\end{equation}
With this we regularize the momentum via the corresponding finite difference operator:\begin{equation}
P_\mathcal{A}\approx \frac{\hbar \sqrt{\mathcal{A} }}{6\sqrt{\pi}\Delta_\mathcal{V}} \frac{1}{2i} \left( e^{i \frac{6\sqrt{\pi}\Delta_\mathcal{V}}{\hbar \sqrt{\mathcal{A}} }P_\mathcal{A}} - e^{-i \frac{6\sqrt{\pi}\Delta_\mathcal{V}}{\hbar \sqrt{\mathcal{A}} }P_\mathcal{A}} \right)
\end{equation}
We define:
\begin{equation}
\bar{p} = \frac{\hbar}{6\sqrt{\pi}\Delta_\mathcal{V}}
\end{equation}
In geometrical units then the dimension of $\bar{p}$ is $L^{-1}$, since $\hbar$ is the Planck length squared.\\
Defining the volume of a sphere of Planck radius as:
\begin{equation}
\mathcal{V}_p = \frac{4\pi}{3}\ell_p^3
\end{equation}
We could rewrite the minimal volume element as:
\begin{equation}
\Delta_\mathcal{V} = \gamma \mathcal{V}_p 
\end{equation}
With $\gamma >0$ a proportionality constant.\\
Since in geometrical units $\hbar = \ell_p^2$, we get:
\begin{equation}
\label{eqn:pbargamma}
\bar{p} = \frac{\ell_p^2}{6\sqrt{\pi} \gamma \mathcal{V}_p} = \frac{1}{ 8\pi^{\frac{3}{2}}\ell_p}\frac{1}{\gamma}
\end{equation}
We use this to get the modified momentum as:
\begin{equation}
P_\mathcal{A}^2 \approx \mathcal{A}\;\bar{p}^2\sin ^2\left(\frac{ P_{\mathcal{A}}}{\bar{p}\sqrt{\mathcal{A}} }\right)
\end{equation}
Substituting this in the Hamiltonian (\ref{eqn:ham_SQ}) we get the effective Hamiltonian density:
\begin{equation}
\label{eqn:poly_hamilt_old}
\mathcal{H}_{pol} = -\frac{\Xi}{2}  \left(32 \pi ^{3/2} \mathcal{A}^{3/2} \bar{p}^2 \sin ^2\left(\frac{ P_{\mathcal{A}}}{\bar{p}\sqrt{\mathcal{A}} }\right)+\frac{ \kappa _S}{\Xi^2}\frac{\sqrt{\mathcal{A}} }{ \sqrt{4\pi }}- 2\frac{ P_\tau}{\Xi }\right)
\end{equation}
Defining for increased readability:
\begin{equation}
\kappa _P = 64 \pi ^2 \Xi ^2 \bar{p}^2
\end{equation}
The effective Hamiltonian becomes:
\begin{equation}
\label{eqn:poly_hamilt}
\mathcal{H}_{pol} = -\frac{\Xi}{2}  \left(\frac{\kappa _P }{\Xi^2}\frac{\mathcal{A}\sqrt{\mathcal{A}}}{\sqrt{4\pi}} \sin ^2\left(\frac{ P_{\mathcal{A}}}{\bar{p}\sqrt{\mathcal{A}} }\right)+\frac{ \kappa _S}{\Xi^2}\frac{\sqrt{\mathcal{A}} }{ \sqrt{4\pi }}- 2\frac{ P_\tau}{\Xi }\right)
\end{equation}
The equations for the dust proper time and its conjugate momentum are unchanged, so their solutions remain the same as before. The equations for the momentum and area are instead modified by the new effective theory, and are given by:
\begin{align}
			\dot{\mathcal{A}} &=  -\frac{\kappa _P\mathcal{A}  \sin \left(\frac{P_{\mathcal{A}}}{ \bar{p}\sqrt{\mathcal{A}} }\right) \cos \left(\frac{P_{\mathcal{A}}}{ \bar{p}\sqrt{\mathcal{A}}}\right)}{  \bar{p}\sqrt{4\pi } \Xi } \label{eqn:A_pol_eom}\\
			\dot{P}_{\mathcal{A}}  &=   \frac{\kappa_S }{8\Xi    \sqrt{\pi\mathcal{A}}}  + \frac{3 \kappa _P \mathcal{A}}{8 \Xi\sqrt{\pi\mathcal{A}}}\sin ^2\left(\frac{P_{\mathcal{A}}}{\bar{p}\sqrt{\mathcal{A}} }\right) -\frac{\kappa _P }{8\Xi  \sqrt{\pi } }\frac{P_{\mathcal{A}}}{\bar{p} } \sin \left(\frac{2 P_{\mathcal{A}}}{\bar{p}\sqrt{\mathcal{A}} }\right)\label{eqn:PA_pol_eom}
\end{align}
\begin{figure}[ht]
	\centering
  \includegraphics[scale=0.75]{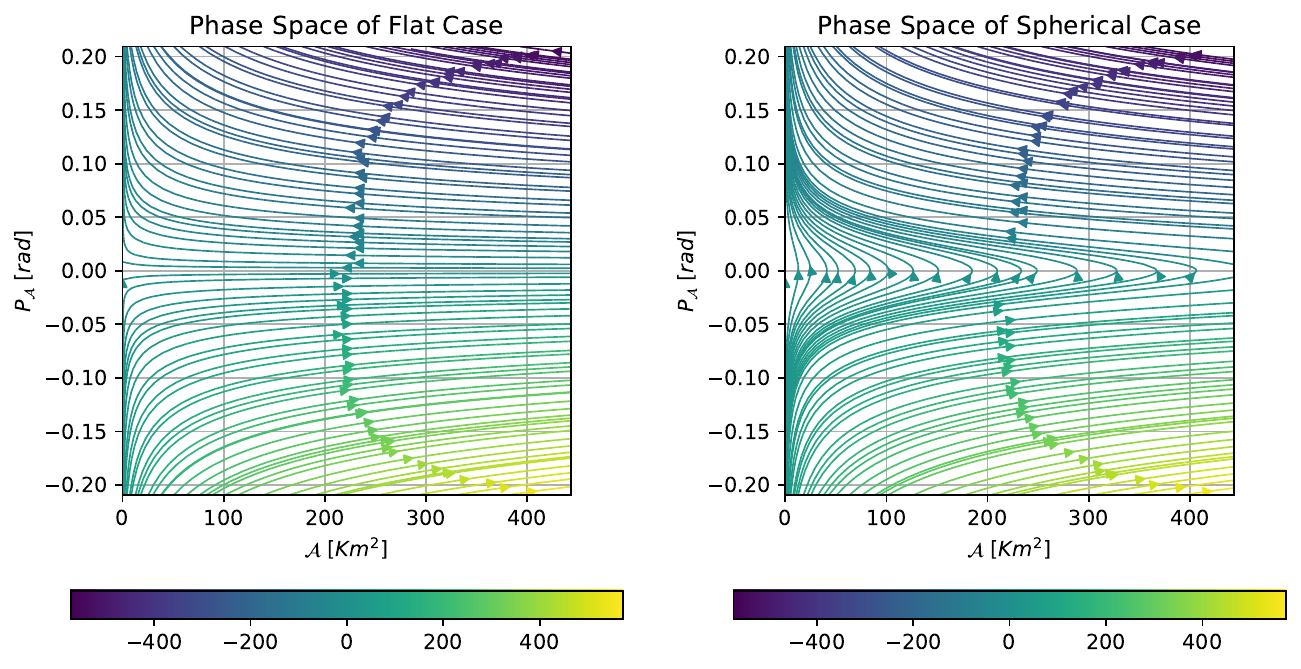}
  \caption{Phase Portraits in Geometrical Units for both the Flat and Spherical models.}
  \label{fig:phase_prt}
  \end{figure}
In Figure \ref{fig:phase_prt} the phase portrait of both models is shown. It is evident that, for a fixed value of $\mathcal{A}$, that later we will see to be the minimum value for this canonical variable, the momentum will switch discontinuously from $P_\mathcal{A}$ to $-P_\mathcal{A}$: this behaviour will be analyzed later in subsection \ref{subsec:mome} when studying the momentum dynamics.\\
We will refer to the positive momentum region as the in-falling region, and to the negative momentum region as the out-going region.\\
Now we need to study the area dynamics in order to understand what is going on during the collapse.
\subsection{\label{subsec:area}Area Dynamics}
\begin{figure}[ht]
	\begin{minipage}{0.48\textwidth}
		\centering
	  \includegraphics[scale=0.55]{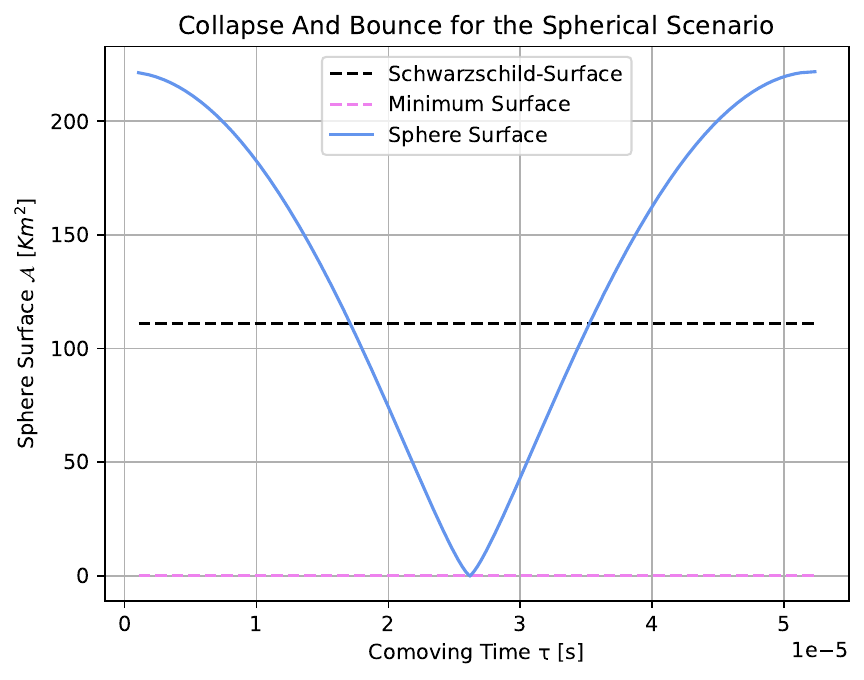}
  \caption{Collapse and Bounce of the Dust Sphere in the Spherical case. The minimum area is $\mathcal{A}_{min} = 1.61\cdot 10^{-49} \textit{Km}^2$}
  \label{fig:sph_surf}
	\end{minipage}
	\hfill
	\begin{minipage}{0.48\textwidth}	
	\centering
  \includegraphics[scale=0.55]{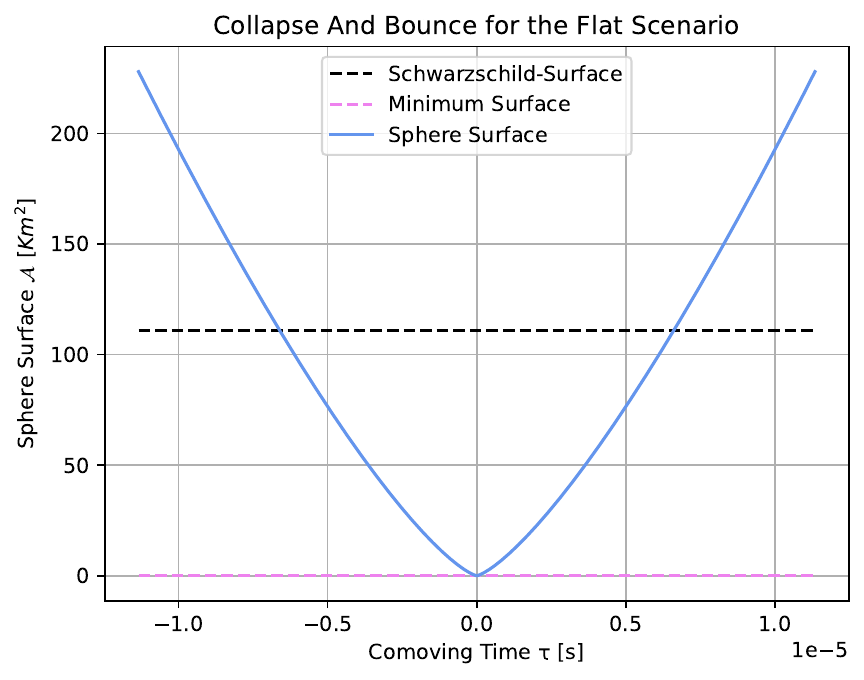}
  \caption{Collapse and Bounce of the Dust Sphere in the Flat case. The minimum area is $\mathcal{A}_{min} = 1.06\cdot 10^{-49} \textit{Km}^2$}
  \label{fig:flt_surf}
  	\end{minipage}
  \end{figure}

The first thing needed in order to study the area dynamics is to remove the momentum dependence from the equation of motion. In order to do so we make use of the Hamiltonian constraint $\mathcal{H}_{pol} =0$ to express the sine function:
\begin{equation}
\label{eqn:constr_pol}
 \sin ^2\left(\frac{ P_{\mathcal{A}}}{\bar{p}\sqrt{\mathcal{A}} }\right)=\frac{4M_S \sqrt{\pi\mathcal{A} }-\kappa _S\mathcal{A} }{\kappa _P\mathcal{A}^{2}}
\end{equation}
This means that:
\begin{equation}
\label{eqn:constr_pol_cos}
\cos^2 \left(\frac{P_{\mathcal{A}}}{ \bar{p}\sqrt{\mathcal{A}}}\right) = 1- \frac{4M_S \sqrt{\pi\mathcal{A} }-\kappa _S\mathcal{A} }{\kappa _P\mathcal{A}^{2}}
\end{equation}
We can substitute those into:
\begin{equation}
\label{eqn:vsqrd1}
\dot{\mathcal{A}}^2 = \frac{\kappa _P^2\mathcal{A}^2  \sin^2 \left(\frac{P_{\mathcal{A}}}{ \bar{p}\sqrt{\mathcal{A}} }\right) \cos^2 \left(\frac{P_{\mathcal{A}}}{ \bar{p}\sqrt{\mathcal{A}}}\right)}{  4\pi\Xi^2  \bar{p}^2 }
\end{equation}
To obtain:
\begin{equation}
\label{eqn:vsqrd2}
\dot{\mathcal{A}}^2 = \frac{\kappa _P}{  4\pi\Xi^2  \bar{p}^2 } \left(4M_S \sqrt{\pi\mathcal{A} }-\kappa _S\mathcal{A}\right) \left(1-\frac{4M_S \sqrt{\pi\mathcal{A} }-\kappa _S\mathcal{A} }{\kappa _P\mathcal{A}^{2}}\right)
\end{equation}
And, taking the square roots, the velocity can be expressed as:
\begin{equation}
\label{eqn:area_pol_eom_v2}
\dot{\mathcal{A}}^\pm = \pm\frac{1}{  \bar{p}\sqrt{4\pi } \Xi }\sqrt{\kappa _P\left(4M_S \sqrt{\pi\mathcal{A} }-\kappa _S\mathcal{A}\right) \left(1-\frac{4M_S \sqrt{\pi\mathcal{A} }-\kappa _S\mathcal{A} }{\kappa _P\mathcal{A}^{2}}\right)}
\end{equation}
The minus is for the in-falling sphere, i.e. the collapsing ball of dust, while the plus is for the out-going, i.e. expanding, sphere.\\
Imposing $\dot{\mathcal{A}}^2=0$ will yield the inversion points of the collapsing sphere: those will be the minima and maxima that the surface reaches.\\
We have two sets of solutions, one pair for the spherical case and one for the flat case.\\
Recalling that the area of a sphere with radius the Schwarzschild radius is $\mathcal{A}_{SC}  = 16\pi M_S^2$ we get for the spherical case:
\begin{equation}
\label{eqn:min_spher}
\mathcal{A}_{max} = \frac{\mathcal{A}_{SC} }{\kappa _S^2} \qquad \mathcal{A}_{min} =\frac{1}{ \kappa _P} \frac{\left(\mathfrak{q}-\kappa _S\right)^2}{3 \mathfrak{q}}
\end{equation}
Where we defined, for an increased readability:
\begin{equation}
\mathfrak{q} = \sqrt[3]{\frac{27}{2}\kappa _P\mathcal{A}_{SC} +3\sqrt{3\kappa _P\mathcal{A}_{SC} \left(\frac{27}{4}\kappa _P\mathcal{A}_{SC} +\kappa _S^3\right)}+ \kappa _S^3}
\end{equation}
The flat case gives instead:
\begin{equation}
\label{eqn:min_flt}
\mathcal{A}_{max} =  +\infty \qquad \mathcal{A}_{min} = \sqrt[3]{\frac{\mathcal{A}_{SC} }{\kappa _P^2} } 
\end{equation}
This result shows that the collapsing sphere reaches a minimum that, with respect to the classical case, is not singular anymore, and that with respect to our previous work, is always confined at the Planckian scale!\\
We can now proceed to solve the equation of motion to study the collapsing dynamics of the dust ball.\\
The differential equation one wish to solve is a separable one, given by:
\begin{equation}
\label{eqn:area_ode}
\frac{d\mathcal{A}^\pm}{\sqrt{\kappa _P\left(4M_S \sqrt{\pi\mathcal{A} }-\kappa _S\mathcal{A}\right) \left(1-\frac{4M_S \sqrt{\pi\mathcal{A} }-\kappa _S\mathcal{A} }{\kappa _P\mathcal{A}^{2}}\right)}} = \pm\frac{dt}{  \bar{p}\sqrt{4\pi } \Xi }
\end{equation}
This is easily solvable in the flat case, with the boundary condition that the in-falling and out-going solutions are matched on the minimum:
\begin{equation}
\mathcal{A}^-(t_{min}) = \mathcal{A}^+(t_{min})  = \mathcal{A}_{min}
\end{equation}
Choosing $t_{min} = 0$ yields the flat solution:
\begin{equation}
\mathcal{A}(t)  =  \sqrt[3]{\mathcal{A}_{SC} \left(9\pi t^2 +\frac{1}{\kappa _P}\right)^2}
\end{equation}
The spherical case is not solvable analytically, so one must resort to computational methods.\\
The boundary conditions used are given by:
\begin{equation}
\mathcal{A}^-(t=0) = \mathcal{A}_0 \qquad    \mathcal{A}^-(t_{min}) = \mathcal{A}^+(t_{min})  =  \mathcal{A}_{min}
\end{equation}
With those conditions the trajectory is numerically evaluated and shown in figure (\ref{fig:sph_surf}), alongside with the trajectory for the flat case in figure (\ref{fig:flt_surf}).\\
The parameter chosen are:
\begin{equation}
\kappa_S = \frac{\sqrt{2}}{2} \qquad M_S = M_\odot \qquad\gamma=2 \;\;\rightarrow\;\;\bar{p} = \frac{1}{16\pi^{\frac{3}{2}}\ell_p}
\end{equation} 

Defining the density of the dust ball as:
 \begin{equation}
 \rho = \frac{M_S}{\mathcal{V}}=\frac{M_S 6\sqrt{\pi}}{\mathcal{A}^{3/2}}
 \end{equation}
 And substituting the minima found in equations (\ref{eqn:min_spher} - \ref{eqn:min_flt}), we see that at the bounce the density is of order $\kappa_P \approx 1/\ell_p^2$, hence Planckian.\\
This shows that also the $\bar{\mu}$ scheme predicts the bouncing of the dust sphere and the subsequent transition from a black hole to a white hole, but now this always happens in the Planck regime, where quantum effects are important.\\
Now we turn our attention to the momentum equation.
\subsection{\label{subsec:mome}Momentum Dynamics}
\begin{figure}[ht]
	\begin{minipage}{0.48\textwidth}
		\centering
	  \includegraphics[scale=0.55]{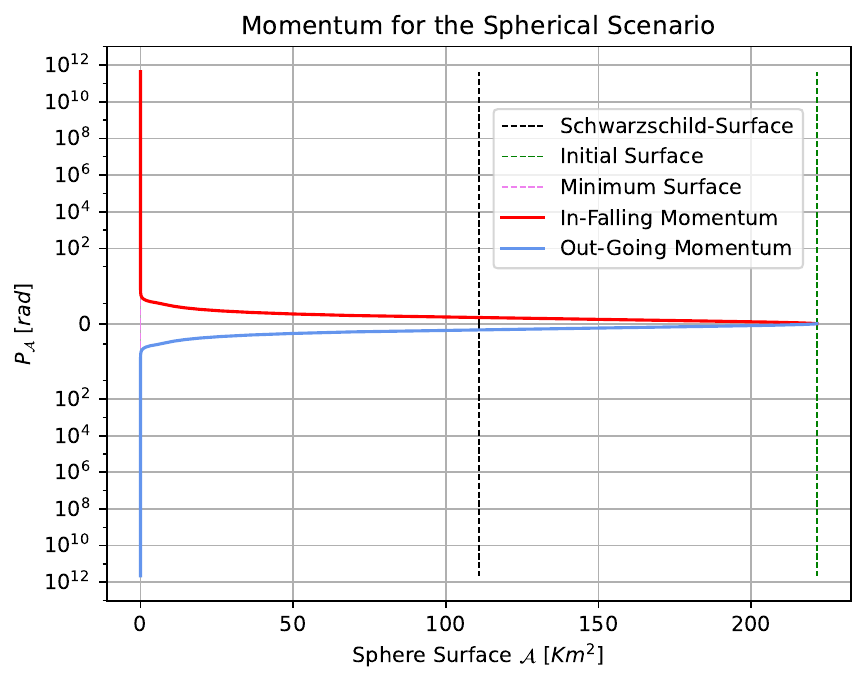}
  \caption{Plot of the Momentum Trajectory against Surface in the Spherical model for both the In-Falling and Out-Going branches.\\A Symmetric-Log scale is used to dispaly both branches. The y-axis displays only the magnitude of the momentum, being the Log-scale unsigned.}
  \label{fig:sph_mom}
	\end{minipage}
	\hfill
	\begin{minipage}{0.48\textwidth}	
	\centering
  \includegraphics[scale=0.55]{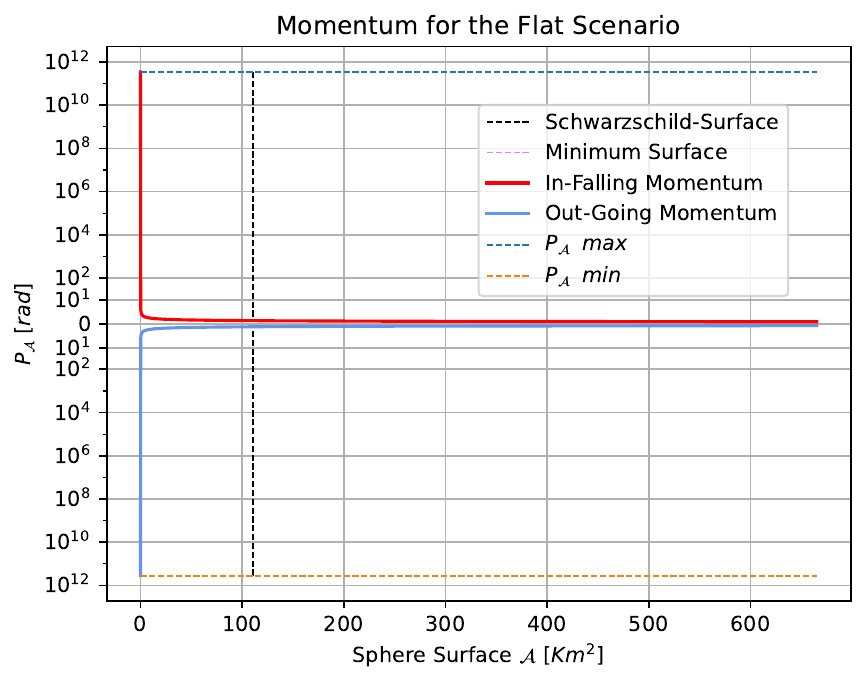}
  \caption{Plot of the Momentum Trajectory against Surface in the Flat model for both the In-Falling and Out-Going branches.\\A Symmetric-Log scale is used to dispaly both branches. The y-axis displays only the magnitude of the momentum, being the Log-scale unsigned.}
  \label{fig:flt_mom}
  	\end{minipage}
  \end{figure}

Now we must solve the equation for the momentum variable.\\
We are more interested in the behaviour of the momentum as the surface changes rather than the behaviour of momentum in time, so we aim to find an expression for $P_\mathcal{A}(\mathcal{A})$.\\
To do so we substitute the constraints given by equations (\ref{eqn:constr_pol}) and (\ref{eqn:constr_pol_cos}), in the same way as we did with $\dot{\mathcal{A}}^\pm$.\\
First we obtain:
\begin{align}
\label{eqn:rdotP}
\dot{\mathcal{A}}^\pm &= \pm\frac{1}{  \bar{p}\sqrt{4\pi } \Xi }\sqrt{\kappa _P\left(4M_S \sqrt{\pi\mathcal{A} }-\kappa _S\mathcal{A}\right) \left(1-\frac{4M_S \sqrt{\pi\mathcal{A} }-\kappa _S\mathcal{A} }{\kappa _P\mathcal{A}^{2}}\right)} \\
\label{eqn:PdotP}
\dot{P}_\mathcal{A}^\pm  &= \frac{\bar{p} \left(6 \sqrt{\pi }M_S-\sqrt{\mathcal{A}} \kappa _S\right) \pm P_\mathcal{A} \sqrt{-\frac{\left(\sqrt{\mathcal{A}} \kappa _S-4 \sqrt{\pi }M_S\right) \left(\sqrt{\mathcal{A}} \left(\mathcal{A}\kappa _P+\kappa _S\right)-4 \sqrt{\pi } M_S\right)}{\mathcal{A}}}}{ \bar{p}\sqrt{16\pi }  \Xi  \mathcal{A}}
\end{align}
Now in order to obtain  $P_\mathcal{A}(\mathcal{A})$ we obtain the derivative of $P_\mathcal{A}$ with respect to $\mathcal{A}$: integrating this will give us the desired function.\\
Since:
\begin{equation}
 P_\mathcal{A}' =\frac{\partial P_\mathcal{A}}{\partial \mathcal{A}} = \frac{\dot{P}_\mathcal{A}}{\dot{\mathcal{A}}}
\end{equation}
Using the expressions (\ref{eqn:rdotP}) and (\ref{eqn:PdotP}) one obtains:
\begin{equation}
\label{eqn:pofa}
\frac{dP_\mathcal{A}^\pm}{d\mathcal{A}} = \frac{P_\mathcal{A}}{2 \mathcal{A}} \pm \frac{\bar{p} \left(6 \sqrt{\pi } M_S-\sqrt{\mathcal{A}} \kappa _S\right)}{2 \sqrt{-\mathcal{A}\left(\sqrt{\mathcal{A}} \kappa _S-4 \sqrt{\pi } M_S\right) \left(\sqrt{\mathcal{A}} \left(\mathcal{A} \kappa _P+\kappa _S\right)-4 \sqrt{\pi } M_S\right)}}
\end{equation}
The differential equation is easily solvable for the flat case, yielding:
\begin{align}
P^-_\mathcal{A}(\mathcal{A}) =& \sqrt{\mathcal{A}} \left[C_- -\bar{p} \cot ^{-1}\left(\sqrt{\frac{4\sqrt{\pi }M_S}{\mathcal{A}^{3/2} \kappa _P-4 \sqrt{\pi }M_S}}\right)\right]\\
P^+_\mathcal{A}(\mathcal{A}) =& \sqrt{\mathcal{A}} \left[C_+ +\bar{p} \cot ^{-1}\left(\sqrt{\frac{4\sqrt{\pi }M_S}{\mathcal{A}^{3/2} \kappa _P-4 \sqrt{\pi }M_S}}\right)\right]
\end{align}
Imposing the boundary condition that at the start of the collapse, at infinity, the momentum is zero:
\begin{equation}
 P^-_\mathcal{A}(\mathcal{A}_0)= P^+_\mathcal{A}(\mathcal{A}_0) = 0
\end{equation}
 We obtain:
\begin{equation}
C_- = \frac{\pi  \bar{p}}{2} \qquad C_+ = -\frac{\pi  \bar{p}}{2} 
\end{equation}
Finally obtaining:
\begin{align}
\label{eqn:PflatIn}
P^-_\mathcal{A}(\mathcal{A}) =& \;+\bar{p}\sqrt{\mathcal{A}}\arctan\left(\sqrt{\frac{4\sqrt{\pi }M_S}{\mathcal{A}^{3/2} \kappa _P-4 \sqrt{\pi }M_S}}\right)\\
\label{eqn:PflatOut}
P^+_\mathcal{A}(\mathcal{A}) =& \;-\bar{p}\sqrt{\mathcal{A}}\arctan\left(\sqrt{\frac{4\sqrt{\pi }M_S}{\mathcal{A}^{3/2} \kappa _P-4 \sqrt{\pi }M_S}}\right)
\end{align}
On the minimum surface the momentum has opposite behaviour, since:
\begin{equation}
P^-_\mathcal{A}(\mathcal{A}_{min}) = \frac{\bar{p}\pi\sqrt{\mathcal{A}_{min}}}{2} \qquad P^+_\mathcal{A}(\mathcal{A}_{min}) = -\frac{\bar{p}\pi\sqrt{\mathcal{A}_{min}}}{2}
\end{equation}
This shows that when the dust sphere bounces there is a discontinuity in the momentum, analogous to an elastic bounce against a potential wall, where $p\rightarrow -p$ when the bounce occurs.\\
The differential equation (\ref{eqn:pofa}) in the spherical case is not analytically solvable, so must be solved numerically, just like the equation for the surface.\\
The boundary conditions imposed are the same as the flat case, namely:
\begin{equation}
 P^-_\mathcal{A}(\mathcal{A}_0)= P^+_\mathcal{A}(\mathcal{A}_0) = 0
\end{equation}
In this case $\mathcal{A}_0$ is finite and defined by the parameter $\kappa_S$, as already stated.\\
The momentum trajectory against the surface is numerically evaluated and shown in figure (\ref{fig:sph_mom}), alongside with the solutions for the flat case in figure (\ref{fig:flt_mom}).\\
The parameter chosen are the same as the ones for the surface trajectory plots:
\begin{equation}
\kappa_S = \frac{\sqrt{2}}{2} \qquad M_S = M_\odot \qquad\gamma=2 \;\;\rightarrow\;\;\bar{p} = \frac{1}{16\pi^{\frac{3}{2}}\ell_p}
\end{equation} 
The same discontinuity encountered in the flat case is also present in the spherical model, as we expected from the phase portraits shown in section \ref{sec:effective_model}, showing that this inversion behaviour is a characteristic intrinsic to the $\bar{\mu}$ effective model, independently of the initial conditions chosen for the collapse.\\
This discontinuity, together with the inversion of the velocity sign, is what characterize a transition from a state representing a black hole to a state representing a white hole: the transition is not smooth and happens abruptly and immediately, once the surface has reached a specified value, the minimum in this case.\\
More precisely, the quantities characterizing the black hole state, the sign of its momentum and velocity, are reversed after the bounce happens, so that after reaching the minimal surface, its motion is equivalent to the time reversal of the collapsing ball: in this sense we obtained a white hole solution, intended as a time-reversed black hole one.
We now continue analyzing more phenomenological aspects related to this effective model in section \ref{sec:phenom}, in order to better understand what this model physically encompass.
\section{\label{sec:phenom}PHENOMENOLOGY}
In this section our goal is to study the phenomenological aspects of the model.\\
We will firstly check whether or not the bounce that occurs during the gravitational collapse could happen outside the event horizon or if it is confined inside the horizon.\\
Then we'll show that the polymer corrections can be understood as a quantum-gravity induced pressure, and an expression for the pressure will be derived.
\subsection{\label{subsec:mini}Minima and Consistency Conditions}
\begin{figure}[ht]
	\centering
  \includegraphics[scale=0.75]{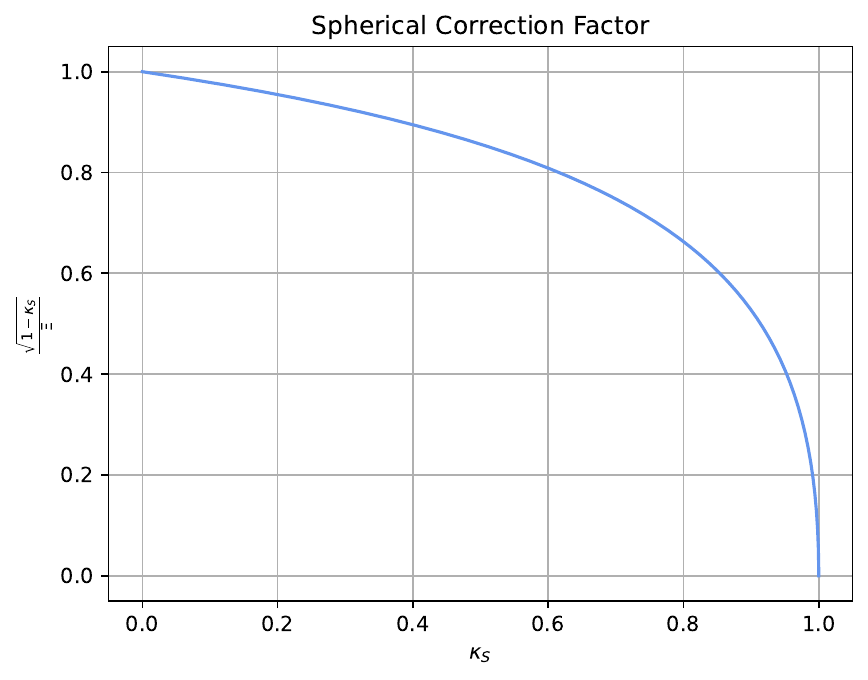}
  \caption{Spherical correction factor to the Schwarzschild radius. It reduces to the flat case when $\kappa_S=0$ and goes to zero when the initial radius is equal to the Schwarzschild radius  $(\kappa_S=1)$.}
  \label{fig:sph_corr}
  \end{figure}

The first thing to analyze is the possibility for the dust ball to bounce outside its event horizon. To do so, the condition to impose is:
\begin{equation}
\mathcal{A}_{min} \geq \mathcal{A}_{SC}
\end{equation}
This condition splits in the two different cases. The flat case reads:
\begin{equation}
\sqrt[3]{\frac{\mathcal{A}_{SC} }{\kappa _P^2} }  \geq  \mathcal{A}_{SC}
\end{equation}
That means:
\begin{equation}
\label{eqn:flatout}
    \mathcal{A}_{SC} < \frac{1}{ \kappa _P}
\end{equation}
Recalling the definition of $\bar{p}$ in equation (\ref{eqn:pbargamma}), the condition becomes:
\begin{equation}
\frac{2M_S}{\ell_p}\leq \frac{\gamma}{2}
\end{equation}
For the spherical case the condition is given by:
\begin{equation}
\frac{1}{ \kappa _P} \frac{\left(\mathfrak{q}-\kappa _S\right)^2}{3 \mathfrak{q}} \geq  \mathcal{A}_{SC}
\end{equation}
This translates into the condition:
\begin{equation}
\frac{2M_S}{\ell_p}\leq \frac{\gamma}{2}\frac{\sqrt{1-\kappa_S}}{\Xi}
\end{equation}
Notice that $\frac{\sqrt{1-\kappa_S}}{\Xi}$ is a strictly decreasing function of the parameter $0 < \kappa_S \leq 1$, hence it further reduces, with respect to the flat case, the allowed values for $2M_S$ admissible to bounce outside the event horizon, as shown in figure (\ref{fig:sph_corr}).\\
The conditions clearly show that the bounce happens outside the event horizon only for dust balls that possess a Planckian Schwarzschild radius, given that $\gamma$ is of the order of unity, namely that $\Delta_\mathcal{V}$ goes approximately as $\ell_p^3$, aside some small multiplicative factors.\\
\subsection{\label{subsec:pressure}The Quantum Gravitational Pressure}
We will know show that the origin of the quantum effects can be understood as coming from a quantum gravitational pressure. This means that our effective theory can be understood as the addition of a pressure term in the stress energy tensor of the classical equations, with this new pressure originating at the quantum level from quantum gravity effects.\\
The starting point is to recall the Friedmann equations for a homogeneous isotropic dust:
\begin{equation}
\frac{\ddot{a}}{a} = -\frac{4\pi}{3}\left(\rho + 3p\right)
\end{equation}
We can rewrite this equation as:
\begin{equation}
\frac{\ddot{R}_s}{R_s} = -\frac{4\pi}{3}\rho -4\pi p
\end{equation}
In area variables it reads:
\begin{equation}
\label{eqn:fried_acc}
\frac{\ddot{R}_s}{R_s} = \frac{\ddot{\mathcal{A}}}{2\mathcal{A}} - \frac{\dot{\mathcal{A}}^2}{4\mathcal{A}^2}
\end{equation}
Since the dust is pressureless, the classical equation is given by:
\begin{equation}
\left.\frac{\ddot{R}_s}{R_s}\right|_{CL} = -\frac{8\pi^{3/2}M_S}{\mathcal{A}^{3/2}} = -\frac{4\pi}{3}\rho
\end{equation}
To find the pressure the idea is to subtract $-\frac{4\pi}{3}\rho$ from the polymer $\left.\frac{\ddot{R}_s}{R_s}\right|_{POL}$, so that the remaining terms are equal to $-4\pi \mathcal{P}_{pol}$, where $ \mathcal{P}_{pol}$ is the pressure generated by the polymer effects, a quantum gravity-induced pressure.\\
We need first to evaluate $\ddot{\mathcal{A}}$, that in the effective theory is given by:
\begin{equation}
\ddot{\mathcal{A}} = \dot{\mathcal{A}}\frac{\partial  \dot{\mathcal{A}}}{\partial \mathcal{A}} +  \dot{P}_\mathcal{A}\frac{\partial  \dot{\mathcal{A}}}{\partial P_\mathcal{A}}
\end{equation}
This yields:
\begin{equation}
\ddot{\mathcal{A}} = 4\pi \left[-\kappa _P\mathcal{A}  \sin ^2\left(\frac{ P_{\mathcal{A}}}{\bar{p}\sqrt{\mathcal{A}} }\right) \left(\cos \left(\frac{ 2P_{\mathcal{A}}}{\bar{p}\sqrt{\mathcal{A}} }\right)-2\right)-\kappa _S \cos \left(\frac{ 2P_{\mathcal{A}}}{\bar{p}\sqrt{\mathcal{A}} }\right)\right]
\end{equation}
Inserting this into equation (\ref{eqn:fried_acc}) gives:
\begin{equation}
\left.\frac{\ddot{R}_s}{R_s}\right|_{POL} = \frac{\pi \left(\kappa _P \mathcal{A} \cos \left(\frac{ 4P_{\mathcal{A}}}{\bar{p}\sqrt{\mathcal{A}} }\right)-\left(3 \mathcal{A} \kappa _P+2 \kappa _S\right) \cos \left(\frac{ 2P_{\mathcal{A}}}{\bar{p}\sqrt{\mathcal{A}} }\right)+2 \mathcal{A} \kappa _P\right)}{\mathcal{A}}
\end{equation}
Substituting the constraint (\ref{eqn:constr_pol}) in the trigonometric functions finally gives the Friedmann polymer equations as a function of only the area variable:
\begin{equation}
\left.\frac{\ddot{R}_s}{R_s}\right|_{POL} =-\frac{8\pi^{3/2}M_S}{\mathcal{A}^{3/2}}+\frac{-48 \pi ^{3/2} \sqrt{\mathcal{A}}M_S\kappa _S+4 \pi  \mathcal{A} \kappa _S^2+128 \pi ^2 M_S^2}{\kappa _P\mathcal{A}^3 }
\end{equation}
This means that:
\begin{equation}
\label{eqn:polpr}
\left.\frac{\ddot{R}_s}{R_s}\right|_{POL} - \left.\frac{\ddot{R}_s}{R_s}\right|_{CL} =-4\pi \mathcal{P}_{pol} =  \frac{-48 \pi ^{3/2} \sqrt{\mathcal{A}}M_S\kappa _S+4 \pi  \mathcal{A} \kappa _S^2+128 \pi ^2 M_S^2}{\kappa _P\mathcal{A}^3 }
\end{equation}
Simplifying the above expression, and recalling that $\mathcal{A}_{SC} = 16\pi M_S^2$ we obtain at last the expression for the polymer pressure:
\begin{equation}
\mathcal{P}_{pol}  = -\frac{\mathcal{A}_{SC} }{\mathcal{A}^3 \kappa _P}\left[\left(\sqrt{\frac{\mathcal{A}}{\mathcal{A}_{0}}}-\frac{3}{2}\right)^2-\frac{1}{4}\right]
\end{equation}
The term $\left(\sqrt{\frac{\mathcal{A}}{\mathcal{A}_{0}}}-\frac{3}{2}\right)^2-\frac{1}{4}$ is always positive since $\mathcal{A}<\mathcal{A}_0$ always, hence the pressure is always negative. \\
This means that this pressure is to be understood as a volume tension and not properly as a pressure.\\
Note also that at the start of the collapse, namely $\mathcal{A}=\mathcal{A}_0$ in the spherical case and $\mathcal{A}= \infty$  in the flat one, the pressure vanishes, so the effects starts to kick in as soon as the dust sphere is starting to compress, antagonizing the collapsing gravitational force, in a way that is similar to an elastic spring.\\
The pressure vanishes also in the $\Delta_\mathcal{V}\rightarrow 0$ limit, so it is evident that the polymer effective theory is the source of such tension, as anticipated at the start of this subsection.\\
\textcolor{black}{We'll analyze in the next section what it means to join the effective dynamics of the interior with the classical one that lies outside the collapsing sphere: does it violates classical General Relativity?
\subsection{\label{subsec:junction}The Junction Conditions and Violations of General Relativity}
A key point in defining solutions to General Relativity by merging two separate solutions is that on the separating hypersurface, called $\Sigma$, the four dimensional metric must be continuous. A similar condition must also be imposed on the four dimensional Riemann curvature tensor. However, if such conditions is not met, then spacetime is singular on $\Sigma$. \\
The meaning of this singularity on the hypersurface is to be interpreted as the presence of a thin-shell/surface membrane on $\Sigma$, with a stress-energy tensor $T^{\mu\nu}_\Sigma$ for the membrane.\\
Defining as $\Sigma$ the hypersurface $r=r_S$, we want the metric inside and outside $\Sigma$ to be joined without discontinuities. The interior (FLRW) part of the spacetime manifold will be labeled by $\mathcal{V}^-$, while the exterior is labeled by $\mathcal{V}^+$. We then define coordinates $x^\mu_-$ and $x^\mu_+$ for both, while on $\Sigma$ we define coordinates $y^a$. For the metrics (\ref{eqn:FLRW_ds}) and (\ref{eqn:SCH_ds}) in the comoving gauge those are:
\begin{equation}
    x^\mu_- = (\tau, r, \theta, \phi) \qquad \qquad x^\mu_+ = (T(\tau), R(\tau), \theta, \phi)
\end{equation}
On $\Sigma$ we have instead:
\begin{equation}
    y^a = (\tau, \theta, \phi)
\end{equation}
The projector from the four dimensional space to the hypersurface $\Sigma$ is given by:
\begin{equation}
    ^\pm e^\mu_a = \frac{\partial x^\mu_\pm}{\partial y^a}
\end{equation}
For both the inside and outside we have:
\begin{equation}
    ^\pm e^\mu_\theta = \delta^\mu_\theta \qquad \qquad  ^\pm e^\mu_\phi = \delta^\mu_\phi
\end{equation}
While for $ ^\pm e^\mu_\tau$ we have:
\begin{equation}
   ^- e^\mu_\tau = (1,0, 0,0)= \delta^\mu_\tau \qquad \qquad  ^+ e^\mu_\tau = (\dot{T}(\tau), \dot{R}(\tau), 0,0)
\end{equation}
We also have the normal to the hypersurface $\Sigma$ that must satisfy:
\begin{equation}
    ^\pm n_\mu  ^\pm e^\mu_a = 0 \qquad \forall a
\end{equation}
Since the projectors are tangential to $\Sigma$.\\
Those vectors are:
\begin{equation}
    ^-n_\mu = \bigg(0,\frac{a}{\sqrt{1-\kappa_S}}\;,0,0\bigg) \qquad   ^+n_\mu = \big(-\dot{R},\dot{T},0,0\big)
\end{equation}
With this is possible to define the induced metric on $\Sigma$, that must be continuous when viewed from either side, to ensure a consistent solution.\\
The induced metric is then defined as:
\begin{equation}
    ^\pm h_{ab}= (^\pm e^\mu_a) (^\pm e^\nu_b)g^\pm_{\mu\nu}
\end{equation}
We also define the extrinsic curvature on the hypersurface as:
\begin{equation}
\label{eqn:K_junc}
 ^\pm K_{ab}= (^\pm e^\mu_a) (^\pm e^\nu_b) \nabla_\nu n^\pm_{\mu}
\end{equation}
This is used to define the stress-energy tensor of the membrane.\\
First define, for a tensor living on $\Sigma$:
\begin{equation}
    [A] = \;^+A|_\Sigma - \;^-A|_\Sigma 
\end{equation}
This will encode the continuity of the tensor $A$ across $\Sigma$. The tensor is continuous if $[A] = 0$.\\
Then the stress-energy tensor of the membrane is:
\begin{equation}
\begin{split}
    T^{\mu\nu}_\Sigma &= \delta(r_S) S^{ab}e^\mu_a e^\nu_b\\
    S_{ab} &= -\frac{1}{8\pi}\bigg([K_{ab}] - [K]h_{ab}\bigg)
\end{split}
\end{equation}
Thus the failure of $[K_{ab}]$ to vanish, which is equivalent to the Riemann tensor being discontinuous, implies the existence of a membrane with non-zero stress-energy on $\Sigma$.\\
What we have to do is then impose the condition:
\begin{equation}
    [h_{ab}] = 0
\end{equation}
In order to have a continuous metric, and then check if under such conditions the tensor $S_{ab}$ is vanishing or not. For the two metrics (\ref{eqn:FLRW_ds}) and (\ref{eqn:SCH_ds}), this condition becomes:
\begin{equation}
    \begin{split}
        h_{\tau \tau}:& \qquad -1 = -F(R)\dot{T} + \frac{\dot{R}^2}{1-\kappa_S} +2\dot{T}\dot{R}\sqrt{1 - \frac{F(R)}{1-\kappa_S}}\\
         h_{\theta \theta}:& \qquad  a^2 r_S^2 = R^2\\
          h_{\phi \phi}:& \qquad  a^2 r_S^2 \sin^2\theta = R^2\sin^2\theta
    \end{split}
\end{equation}
Those can be imposed without troubles, and already were imposed from the start, when we said we considered $R_s = R(\tau)$ as the star boundary in comoving time throughout the whole paper.\\
Now, with those conditions explicited, we can go and check whether we have membrane stress-energy tensor or not.\\
We first define the function:
\begin{equation}
    \mathcal{E}(R_s,\dot{R}_s) = \sqrt{\dot{R}_s^2 + F(R_s)} = \sqrt{\dot{R}_s^2 + 1-\frac{2M_S}{R_s}}
\end{equation}
That is the energy of a dust particle, and is the function that will encode any effective dynamics through $\dot{R}_s$.\\
Now using equation (\ref{eqn:K_junc}) we can easily calculate the extrinsic curvatures, without making use of any equation of motion. This results in:
\begin{align}
    \mathcal{V}^-:&\qquad ^-K_{\tau\tau} = 0 \qquad ^-K_{\theta\theta}= ar_S\sqrt{1-\kappa_S}\qquad ^-K_{\phi\phi}=\sin^2\theta\; ar_S\sqrt{1-\kappa_S}\\
    \mathcal{V}^+:&\qquad ^+K_{\tau\tau} = \dot{R}_sa^T - \dot{T}a^R \qquad ^+K_{\theta\theta}= \mathcal{E}(R_s,\dot{R}_s)R_s\qquad ^+K_{\phi\phi}=\mathcal{E}(R_s,\dot{R}_s)R_s\sin^2\theta
\end{align}
Where $a^R$ and $a^T$ are the radial and temporal components of the acceleration four-vector in $ \mathcal{V}^+$. Imposing the metric continuity conditions leads to:
\begin{align}
    \mathcal{V}^-:&\qquad ^-K_{\tau\tau}= 0 \qquad &^-K_{\theta\theta}= R_s\sqrt{1-\kappa_S}\qquad ^-K_{\phi\phi}=\sin^2\theta\;R_s\sqrt{1-\kappa_S}\\
    \mathcal{V}^+:& \qquad ^+K_{\tau\tau} = -\frac{a^R}{\mathcal{E}(R_s,\dot{R}_s)} \qquad &^+K_{\theta\theta}= \mathcal{E}(R_s,\dot{R}_s)R_s\qquad ^+K_{\phi\phi}=\mathcal{E}(R_s,\dot{R}_s)R_s\sin^2\theta
\end{align}
With:
\begin{equation}
        a^R = \ddot{R}_s + \frac{F'}{2} = \ddot{R}_s +\frac{M_S}{R^2_s}
\end{equation}
The $S_{ab}$ tensor built of these is then:
\begin{equation}
    S_{ab} = -\frac{1}{8\pi}\bigg([K_{ab}] +(^+K_{\tau\tau})h_{ab} -\frac{2h_{ab}}{R_s^2}[K_{\theta\theta}] \bigg)
\end{equation}
We thus have only the $S_{\tau\tau}$ component, the $S_{\theta\theta}$ component and the $S_{\phi\phi}$, that is equal in our scenario to $\sin^2\theta \;S_{\theta\theta}$. The only independent components are thus $S_{\tau\tau}$ and $S_{\theta\theta}$.\\
A small calculation yields:
\begin{align}
    S_{\tau\tau} &= \frac{1}{4\pi R_s}\bigg(\sqrt{1-\kappa_S} - \mathcal{E}(R_s,\dot{R}_s)\bigg)\\
    S_{\theta\theta} &= -\frac{R_s}{8\pi }\Bigg(-R_s\frac{\ddot{R}_s + F'/2}{\mathcal{E}(R_s,\dot{R}_s)} + \sqrt{1-\kappa_S} -\mathcal{E}(R_s,\dot{R}_s)\Bigg)\\
    &= -\frac{R_s^2}{8\pi }\Bigg(4\pi S_{\tau\tau}-\frac{\ddot{R}_s + F'/2}{\mathcal{E}(R_s,\dot{R}_s)} \Bigg)
\end{align}
Now we recast those in our area variables, so that:
\begin{equation}
    \dot{R}^2_s = \frac{\dot{\mathcal{A}}^2}{16\pi\mathcal{A}} \qquad F(R_s) = 1-\frac{4M_S\sqrt{\pi\mathcal{A}}}{\mathcal{A}} \qquad \frac{F'}{2}=\frac{4\pi M_S}{\mathcal{A}}
\end{equation}
We start by analyzing the time component.\\
From the equation of motion (\ref{eqn:A_clas_eom}) and the momentum obtained from the classical Hamiltonian, we easily see that in the classical case $\mathcal{E} = \sqrt{1-\kappa_S}$, so that in the classical case $S_{\tau\tau}$ is zero.\\
In the effective theory, however, we must use equation (\ref{eqn:vsqrd2}) for the $\dot{\mathcal{A}}^2$ term in $\mathcal{E}$. \\
Substituting it in:
\begin{equation}
    \mathcal{E}^2 = \frac{\dot{\mathcal{A}}^2}{16\pi\mathcal{A}} + 1-\frac{4M_S\sqrt{\pi\mathcal{A}}}{\mathcal{A}}
\end{equation}
Yields:
\begin{equation}
    \mathcal{E}^2 = 1-\kappa_S - \frac{(4M_S\sqrt{\pi\mathcal{A}} - \kappa_S\mathcal{A})^2}{\kappa_P \mathcal{A}^3}
\end{equation}
The last term encodes the energy difference between the classical and effective theory. It can be rewritten as:
\begin{equation}
    \Delta_P = \frac{\mathcal{A}_{SC}}{\kappa_P \mathcal{A}^2}\Bigg(1-\sqrt{\frac{\mathcal{A}}{\mathcal{A}_0} }\Bigg)^2
\end{equation}
Note that $\Delta_P$ is always a positive quantity.\\
With this now we see that:
\begin{equation}
     S_{\tau\tau} = \frac{1}{2\sqrt{\pi\mathcal{A}}}\bigg(\sqrt{1-\kappa_S}-\sqrt{1-\kappa_S -\Delta_P} \bigg)
\end{equation}
Since $\Delta_P$ is always positive, this means that $S_{\tau\tau}$ is always positive too, assumed that $1-\kappa_S >\Delta_P$ for all the collapse. We'll come to verify that later.\\
Now let's turn to $S_{\theta\theta}$. \\
In the classical theory, $S_{\tau\tau}$ is zero and $\ddot{R}_s = -F'/2$, so also $S_{\theta\theta}$ vanishes in the classical theory. However, in the effective theory, $S_{\tau\tau}$ is not zero and $\ddot{R}$ is different from the classical one. \\
In particular, if we define:
\begin{equation}
    \frac{F'}{2} = -\left.\ddot{R_s}\right|_{CL}
\end{equation}
We see that we get:
\begin{equation}
    S_{\theta\theta}= -\frac{R_s^2}{8\pi }\Bigg(4\pi S_{\tau\tau}-\frac{\left.\ddot{R_s}\right|_{POL} -\left.\ddot{R_s}\right|_{CL}}{\mathcal{E}(R_s,\dot{R}_s)} \Bigg)
\end{equation}
From equation (\ref{eqn:polpr}) we see that:
\begin{equation}
   \left.\ddot{R_s}\right|_{POL} -\left.\ddot{R_s}\right|_{CL} = - 2\sqrt{\pi\mathcal{A}}\;\;\mathcal{P}_{pol}(\mathcal{A}) 
\end{equation}
Notice that:
\begin{equation}
    \mathcal{P}_{pol}(\mathcal{A}) = \frac{d\Delta_P (\mathcal{A})}{d\mathcal{A}}\equiv \Delta_P' (\mathcal{A})
\end{equation}
So we have, in the end:
\begin{equation}
\begin{split}
         S_{\theta\theta}=& -\frac{\mathcal{A}}{8\pi }\Bigg( S_{\tau\tau}+\frac{\sqrt{\mathcal{A}}\;\;\mathcal{P}_{pol}(\mathcal{A}) }{2\sqrt{\pi}\mathcal{E}(R_s,\dot{R}_s)} \Bigg)\\
         =& -\frac{\mathcal{A}}{8\pi }\Bigg( \frac{\sqrt{1-\kappa_S}-\sqrt{1-\kappa_S -\Delta_P}}{2\sqrt{\pi\mathcal{A}}}+\frac{\Delta_P' (\mathcal{A})\sqrt{\mathcal{A}}}{2\sqrt{\pi}\sqrt{1-\kappa_S -\Delta_P}} \Bigg)
\end{split}
\end{equation}
Of course, $S_{\phi\phi} = S_{\theta\theta} \sin^2\theta$. \\
Now we can see that this can be understood as the surface stress-energy tensor of a perfect fluid with pressure, for which the tensor is given by:
\begin{equation}
    S^{ab} = (\sigma + p)u^au^b + p h^{ab}
\end{equation}
Since $u^a= \delta^a_\tau$ and:
\begin{equation}
    h^{ab} = \text{Diag}\bigg( -1,\frac{1}{R_s^2},\frac{1}{R_s^2\sin^2\theta}\bigg)
\end{equation}
We get:
\begin{equation}
    S^{ab} =  \text{Diag}\bigg(\sigma,\frac{p}{R_s^2},\frac{p}{R_s^2\sin^2\theta} \bigg)
\end{equation}
To determine $\sigma$ and $p$ from our computed stress tensor we start by noting that:
\begin{equation}
    S_{\tau\tau}=S^{\tau\tau}\qquad S_{\theta\theta}=S^{\theta\theta}R_s^4 \qquad S_{\phi\phi}=S^{\phi\phi}R_s^4\sin^4\theta
\end{equation}
This means:
\begin{equation}
    S_{\tau\tau} = \sigma \qquad S_{\theta\theta} = p R_s^2 =\frac{p\mathcal{A}}{4\pi}
\end{equation}
So we have a membrane at $r=r_S$ made out of a perfect fluid with density given by:
\begin{equation}
\label{eqn:mem_dens}
    \sigma = S_{\tau\tau} =  \frac{1}{2\sqrt{\pi\mathcal{A}}}\bigg(\sqrt{1-\kappa_S}-\sqrt{1-\kappa_S -\Delta_P} \bigg)
\end{equation}
And pressure given by:
\begin{equation}
\label{eqn:mem_pres}
    p=\frac{4\pi S_{\theta\theta}}{\mathcal{A}} = -\frac{1}{2}\Bigg( \frac{\sqrt{1-\kappa_S}-\sqrt{1-\kappa_S -\Delta_P}}{2\sqrt{\pi\mathcal{A}}}+\frac{\Delta_P' (\mathcal{A})\sqrt{\mathcal{A}}}{2\sqrt{\pi}\sqrt{1-\kappa_S -\Delta_P}} \Bigg)
\end{equation}
We can now start to see if such fluid violates any of the energy conditions\cite{hawking_ellis_1973} for admissible fluids in GR.\\
For a 2+1 dimensional perfect fluid, those are given by:
\begin{align}
    \textit{Weak Energy Condition (WEC): }& \qquad\sigma \geq 0\\
    \textit{Strong Energy Condition (SEC): }& \qquad 2 p \geq 0\\
     \textit{Null Energy Condition (NEC): }& \qquad\sigma + p \geq 0\\
      \textit{Dominant Energy Condition (DEC): }& \qquad\sigma \geq |p| 
\end{align}
We see immediately that if both the WEC and the SEC are satisfied, then the NEC is automatically satisfied.\\
We start by analyzing the WEC: if $1-\kappa_S > \Delta_P$ for all the collapse, then the density given by (\ref{eqn:mem_dens}) is always positive, hence the WEC is satisfied if $1-\kappa_S > \Delta_P$ holds during the collapse. Since this is an existence condition and not a dynamical requirement, we'll treat this after the full analysis of the energy conditions.\\
Now we turn our attention to the SEC. For $2p$ to be positive we need:
\begin{equation}
     \frac{\sqrt{1-\kappa_S}-\sqrt{1-\kappa_S -\Delta_P}}{2\sqrt{\pi\mathcal{A}}}+\frac{\Delta_P' (\mathcal{A})\sqrt{\mathcal{A}}}{2\sqrt{\pi}\sqrt{1-\kappa_S -\Delta_P}}<0
\end{equation}
The first term, being $\sigma$, is never negative. On the other and, since $\Delta_P' (\mathcal{A})\leq 0$ always, the pressure is positive if:
\begin{equation}
    \left|\Delta_P' (\mathcal{A})\right| > \frac{\sqrt{1-\kappa_S}-\sqrt{1-\kappa_S -\Delta_P}}{\mathcal{A}}\sqrt{1-\kappa_S -\Delta_P}
\end{equation}
Substitution of $-\mathcal{P}_{pol}(\mathcal{A})$ into $ \left|\Delta_P' (\mathcal{A})\right|$ and some algebra shows that this is always satisfied if $1-\kappa_S > \Delta_P$. Thus, if the function $\mathcal{E}(\mathcal{A})$ is defined for all the collapse, then also the SEC is satisfied.\\
This implies that also the NEC is satisfied in the domain of existence of $\mathcal{E}(\mathcal{A})$.\\
We are left only with the need to check the DEC.\\
On the domain of existence of $\mathcal{E}(\mathcal{A})$ the pressure is always positive, so we can omit the absolute value and re-express $p$ as:
\begin{equation}
    |p| =  -\frac{\sigma}{2} -\frac{\Delta_P' (\mathcal{A})\sqrt{\mathcal{A}}}{4\sqrt{\pi}\sqrt{1-\kappa_S -\Delta_P}} 
\end{equation}
The DEC thus reads:
\begin{equation}
    \sigma \geq -\frac{\sigma}{2} -\frac{\Delta_P' (\mathcal{A})\sqrt{\mathcal{A}}}{4\sqrt{\pi}\sqrt{1-\kappa_S -\Delta_P}} 
\end{equation}
Rearranging yields:
\begin{equation}
    \frac{3\sigma}{2} +\frac{\Delta_P' (\mathcal{A})\sqrt{\mathcal{A}}}{4\sqrt{\pi}\sqrt{1-\kappa_S -\Delta_P}}  \geq0
\end{equation}
We now substitute equation (\ref{eqn:mem_dens}) for $\sigma$ and simplify, obtaining:
\begin{equation}
  3 \frac{\sqrt{1-\kappa_S}-\sqrt{1-\kappa_S -\Delta_P}}{\mathcal{A}}  +\frac{\Delta_P' (\mathcal{A})}{\sqrt{1-\kappa_S -\Delta_P}}  \geq 0
\end{equation}
Since, again, $\Delta_P'$ is always negative, this becomes:
\begin{equation}
    3\sqrt{1-\kappa_S -\Delta_P} \frac{\sqrt{1-\kappa_S}-\sqrt{1-\kappa_S -\Delta_P}}{\mathcal{A}} \geq \left|\Delta_P' (\mathcal{A})\right|
\end{equation}
This condition is satisfied only if $1-\kappa_S < \Delta_P$ and thus, since it makes the function $\mathcal{E}(\mathcal{A})$ complex and that's not allowed, it is never satisfied during the collapse.\\
So this means that while the WEC, the SEC and the NEC are always satisfied for a well defined $\mathcal{E}(\mathcal{A})$, a well defined $\mathcal{E}(\mathcal{A})$ means that the DEC is never satisfied whenever the other three conditions are satisfied, and viceversa.\\
This is not an unexpected result, since the same violation of the DEC was also found previously in the literature\cite{Hergott_Husain_2026}, and is the key ingredient that allows a dynamical black-to-white transition, strengthening the case presented in the previous section and there already motivated by the discontinuity in the momentum variable.\\
It is now the time to finally analyze the condition $1-\kappa_S \geq \Delta_P$ that is required to enforce all the energy conditions except the DEC.\\
For the flat case this is given by:
\begin{equation}
    1-\frac{\mathcal{A}_{SC}}{\kappa_P\mathcal{A}^2}\geq 0
\end{equation}
Solved for $\mathcal{A}$ gives:
\begin{equation}
    \mathcal{A}\geq \sqrt{\frac{\mathcal{A}_{SC}}{\kappa_P}}\equiv \mathcal{A}_{\mathcal{E}}
\end{equation}
Now notice that we if want the minimum to be greater than the value at which the energy becomes ill-defined, we must impose $\mathcal{A}_{min} > \mathcal{A}_{\mathcal{E}}$. But this is exactly the condition (\ref{eqn:flatout}) for the minimum to be outside of the event horizon: this means that the junction conditions hold and the three energy conditions are satisfied up until the minimum only by dust balls that possess a Planckian Schwarzschild radius. Otherwise, the junction conditions break before reaching the minimal surface.\\
For the spherical case the condition is instead given by:
\begin{equation}
    1-\kappa_S  \geq  \Delta_P
\end{equation}
Unfortunately, contrary to the flat case, this is impossible to solve analitically.\\
We thus resort to numerical solutions. We fix a Planckian value for the parameter $\Delta_\mathcal{V}$ to $10\mathcal{V}_p$, then we study in a $(\mathcal{A};M_S)-$plane the value of:
\begin{equation}
    \mathcal{E}^2 =   1-\kappa_S  -  \Delta_P
\end{equation}
As a function of mass through $\mathcal{A}_{SC}$ and of the area.\\
In figure (\ref{fig:epsi_ks}) we see such plots for different values of $\kappa_S$, equal to $0.1$, $0.5$ and $0.98$ respectively, with the zeros of $\mathcal{E}^2$ in white, a red color for positive $\mathcal{E}^2$ and blue for the negative. As in the flat case, the function becomes ill-defined before the minimal surface, indicated by the green dashed line in the figures.The cases of Planckian mass are shown in figure (\ref{fig:epsiPL_ks}). Here it's evident that as $\kappa_S$ increases, the zeros of $\mathcal{E}^2$ diverge away from the minimal surface, so that we have its zeros before the minima just for small values of $\kappa_S$ in the Planckian mass regime.\\
\begin{figure}[h!]
		\centering
	  \includegraphics[scale=1]{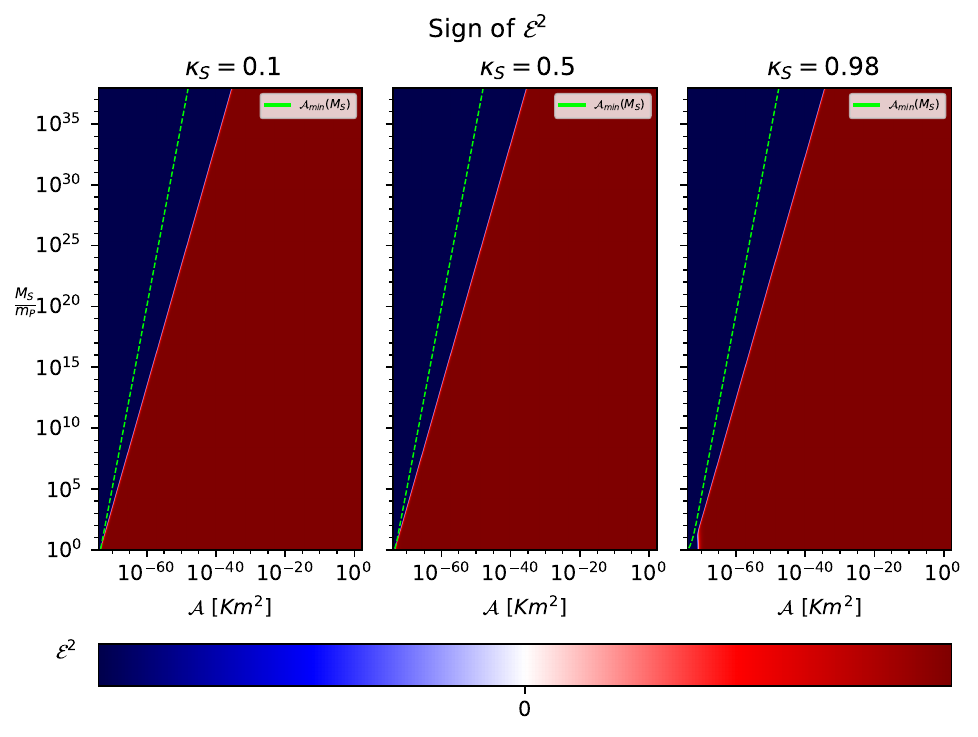}
  \caption{Plot of $\mathcal{E}^2$ with zeros in white, red for positive values and blue for the negative ones. In green, the minimum as a function of $M_S$.}
  \label{fig:epsi_ks}
  \end{figure}\\
  \begin{figure}[h!]
	\centering
  \includegraphics[scale=1]{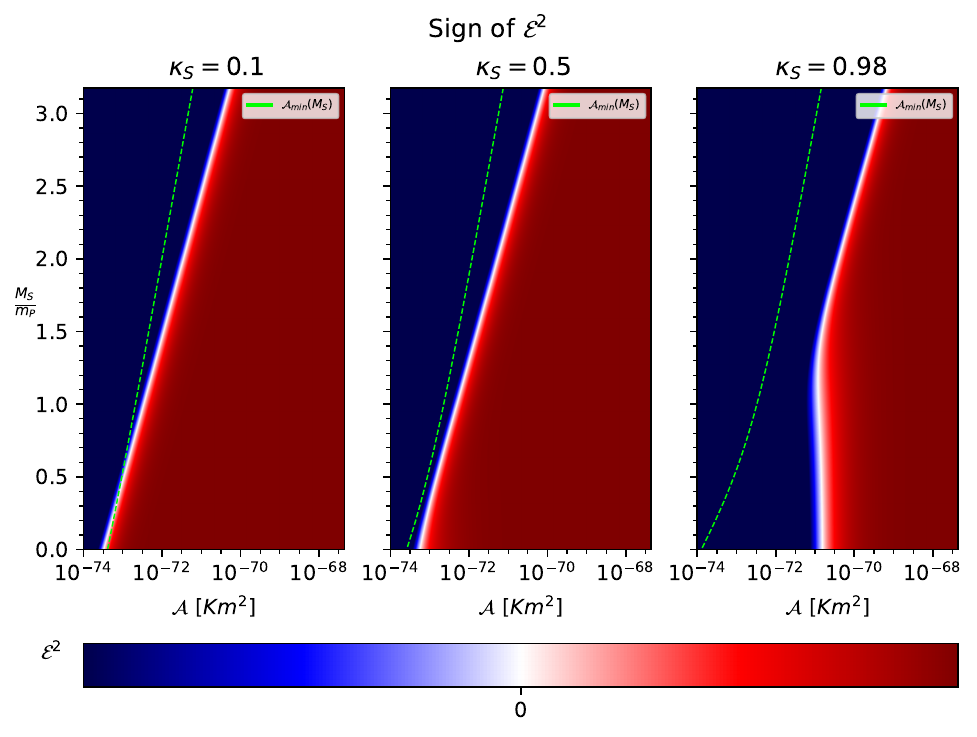}
  \caption{Plot of $\mathcal{E}^2$ for Planckian mass, with zeros in white, red for positive values and blue for the negative ones. In green, the minimum as a function of $M_S$.}
  \label{fig:epsiPL_ks}
  \end{figure}
Before closing we notice that, for non-Planckian masses, even if $\mathcal{A}_{\mathcal{E}}$ is bigger than the minimal surface in both cases, it's still Planckian in magnitude and far deep inside the horizon: this means that the effective description breaks down before reaching the minimal surface, but it does so in the region where we have very strong quantum effects. Here the effective description is not sufficient anymore and rather a full quantum description of the system is instead needed. Those problems are avoided by Planckian objects and low $\kappa_S$, as their entire evolution can be described through the effective framework derived in this manuscript.
}

\section{\label{sec:conc}CONCLUSIONS}
In this paper we studied the possible improvements that the $\bar{\mu}$ scheme polymerization could introduce to the effective model of the Oppenheimer-Snyder gravitational collapse, analyzed in metric variables.
In section \ref{sec:effective_model} we have introduced the new improved model, that with respect to the effective theory previously presented in \cite{Boldorini_2024} yields results more physically clear and expected regarding the quantum effects arising during collapse. The main result of the new model is that the bounce happens, regardless of the specific initial conditions, always at the Planck scale. This behavior is preferred since the Planck scale is where one expects quantum gravity effects to come in.\\
\textcolor{black}{We also confirmed that the bouncing dynamic is a characteristic intrinsic to polymerized effective models, not exclusive to connection variables, and thus potentially useful to implement quantum corrections in other fields where a connection formalism is not largely diffused.}\\
In section \ref{sec:phenom} we studied the more phenomenological aspects of the improved scheme.\\
We have found, similarly to the previous work, that the origin of the repulsive behavior of the polymer effective theory is to be found in a volume tension that counteracts the collapse, trying to restore the initial size of the collapsing sphere. The tension is nevertheless able to alter and stop the collapse only when the dust has reached the Planck scale, enabling for the transition to an out-going solution.\\
We have also found that the minima of the sphere surface are always located inside the event horizon, unless the collapsing star is of Planckian mass, case that allows for the bounce to happen outside the event horizon. This means that, unless the mass of the sphere is Planckian, the bounce always implies a quantum transition from a black hole to a white hole.\\ The pulsating objects theorized in the previous work are then suppressed in this improved model, unless the objects themselves are Planckian in size.\\
This improved $\bar{\mu}$ scheme offers a more viable model to encompass quantum corrections to the classical theory with respect to the previously analyzed formulation, since quantum effects are present only at the Planckian scale.\\
\textcolor{black}{We also analyzed the junction conditions for such a model, where the existence of a surface membrane of stress-energy is found. Such membrane does not violates the energy conditions of GR, except for the dominant energy condition. It is also shown that the effective model breaks down before reaching the minima, as the junction conditions become ill-defined. The model breaks down in the deep quantum regime, signaling the necessity of a full quantum theory to understand the deep quantum regime of gravitational collapse.\\}
Further work will be necessary if one wants to implement, using the $\bar{\mu}$ scheme, more realistic models which include pressure, inhomogeneities or even rotationality.

\section*{Acknowledgments}
We would like to thank Francesco Fazzini for the helpful discussions on the topic.

\printbibliography
\end{document}